\documentclass[journal=jacsat,manuscript=article]{achemso}

\usepackage[version=3]{mhchem} 
\usepackage{color}
\usepackage{siunitx}

\author{Enrico Tapavicza}
\affiliation{Department of Chemistry and Biochemistry, California State University, Long Beach, 1250 Bellflower Boulevard, Long Beach, CA, 90840, USA}
\email{enrico.tapavicza@csulb.edu}
\author{Guido Falk von Rudorff}
\affiliation{University of Vienna, Faculty of Physics, Kolingasse 14-16, AT-1090 Wien, Austria, and
Institute of Physical Chemistry and National Center for Computational Design and Discovery of Novel Materials (MARVEL), Department of Chemistry, University of Basel, Klingelbergstrasse 80, CH-4056 Basel, Switzerland}
\author{David O. De Haan}
\affiliation{Department of Chemistry and Biochemistry, University of San Diego, 5998 Alcala Park, San Diego, CA, 92110, USA}
\author{Mario Contin}
\affiliation{Universidad de Buenos Aires, Facultad de Farmacia y Bioqu\'imica, Departamento de Qu\'imica Analitica y Fisicoqu\'imica, Jun\'in 956, Buenos Aires, C1113AAD, Argentina}
\author{Christian George}
\affiliation{Universit\'e Lyon, Universit\'e Claude Bernard Lyon 1, CNRS, IRCELYON, 69626 Villeurbanne, France}
\author{Matthieu Riva}
\affiliation{Universit\'e Lyon, Universit\'e Claude Bernard Lyon 1, CNRS, IRCELYON, 69626 Villeurbanne, France}
\author{O. Anatole von Lilienfeld}
\affiliation{University of Vienna, Faculty of Physics, Kolingasse 14-16, AT-1090 Wien, Austria, and Institute of Physical Chemistry and National Center for Computational Design and Discovery of Novel Materials (MARVEL), Department of Chemistry, University of Basel, Klingelbergstrasse 80, CH-4056 Basel, Switzerland}

\title[An \textsf{achemso} demo]
{Elucidating atmospheric brown carbon -- Supplanting chemical intuition with exhaustive enumeration and machine learning}
\abbreviations{IR,NMR,UV}

\begin{document}

\begin{tocentry}
\includegraphics[scale=0.044]{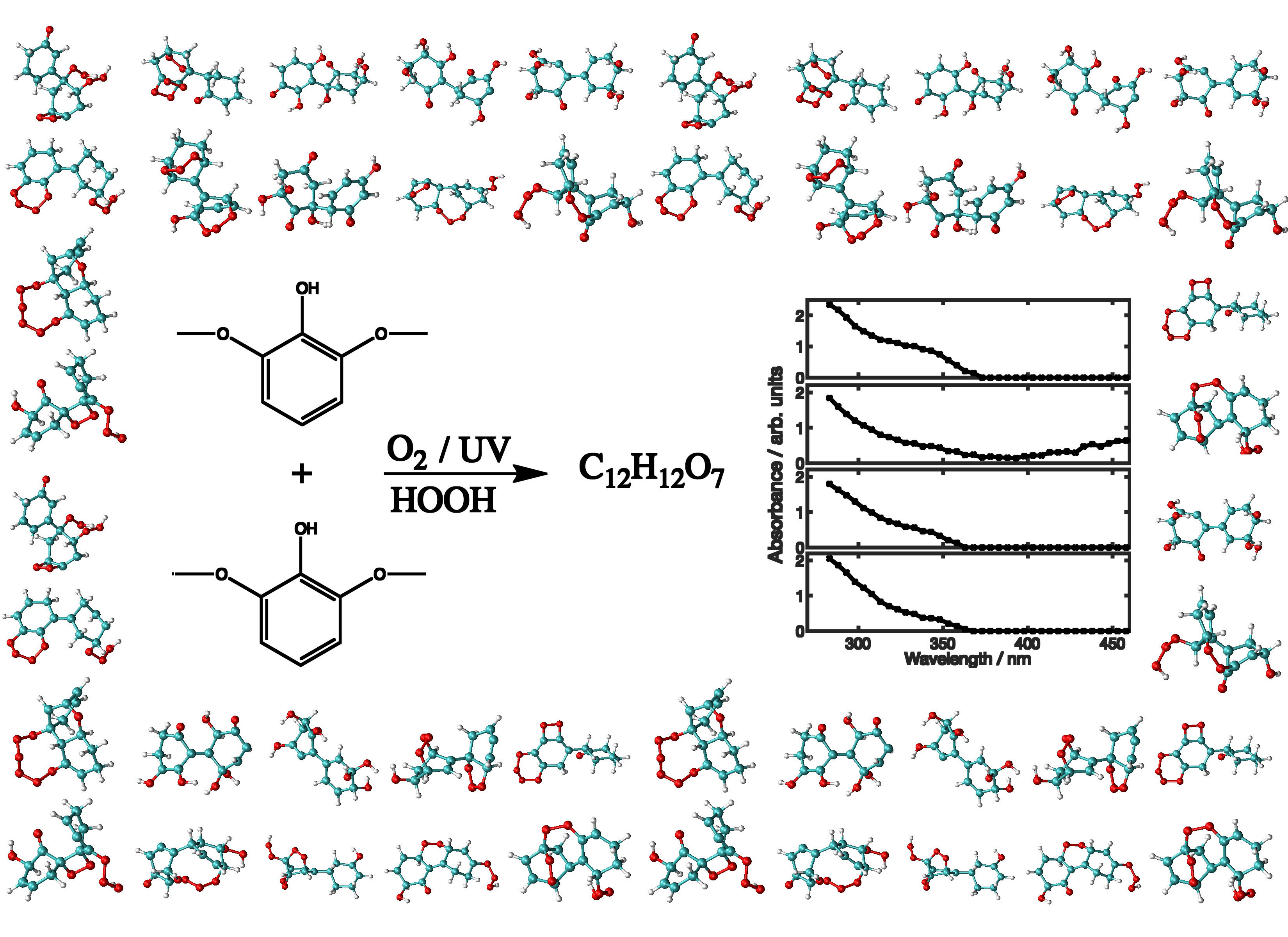}

\end{tocentry}

\begin{abstract}
To unravel the structures of constitutional C$_{12}$H$_{12}$O$_7$ isomers, identified as light-absorbing photooxidation products of syringol in atmospheric chamber experiments, we apply a combined graph-based molecule generator and machine learning workflow.
To accomplish this in a bias-free manner, molecular graphs of the entire chemical subspace of C$_{12}$H$_{12}$O$_7$ were generated,
under only the assumption that the isomers contain two C$_6$-rings; this led to 260 million molecular graphs, of which 120 million were stable structures.
For these structures, using high level quantum chemistry excitation energies and oscillator strengths as training data, we predicted these quantities using kernel ridge regression and simulated UV/Vis absorption spectra.
Then we determined the  probability of the molecules to cause the experimental spectrum within the errors of the different methods. 
Molecules whose spectra were likely to match the experimental spectrum were clustered according to structural features, resulting in clusters of $> 500,000$ molecules.
While we identified several features that correlate with a high probability to cause the experimental spectrum, we show that no clear composition of necessary features can be given. 
Thus, the absorption spectrum is not sufficient to uniquely identify one specific isomer structure.
We show that if more structural features were known from experimental data (especially mass spectrometric fragmentation), the number of structures could be reduced to a few tens of thousands candidates. We offer a procedure to detect when sufficient fragmentation data has been included to reduce the number of possible molecules.
This knowledge could be applied at the stage of molecules generation, as well as at the final filtering stage. Our study suggests that the most efficient strategy to obtain valid candidates is obtained if structural data is applied already at the bias-free molecule generation stage. 
The systematic enumeration, however, is necessary to avoid mis-identification of molecules, while it guarantees that there are no other molecules that would also fit the spectrum in question.

\end{abstract}

\section{Introduction}
Visible light-absorbing secondary organic aerosols (SOAs), also known as brown carbon (BrC), interfere in atmospheric processes and impact climate forcing
\cite{laskin2015chemistry,feng2013brown,Epstein2013,kasthuriarachchi2020light}.
Emerging from biomass burning and from natural and industrial emissions, SOAs constantly undergo several chemical modifications due to reactions in the atmosphere.  
The formation and further reactions of SOA under specific atmospheric conditions can be simulated in atmospheric chamber experiments \cite{denjean2014new}. In these experiments, aerosol-phase reaction products are often extracted from filters and characterized by high-resolution liquid chromatography (LC)/mass spectrometry (MS) analysis with inline UV/Vis absorbance spectroscopy. While the detected mass of the compounds gives complete information about the chemical sum formula, the exact chemical structure remains unknown. 
The characterization of the molecular composition, particularly identifying the major constitutional isomers of SOAs, is a pressing challenge for atmospheric chemistry \cite{schilling2015secondary,de2018nitrogen,fleming2020molecular}.
In chamber experiments, often molecular structures are proposed on the basis of mass spectrometric fragment data, absorption estimates, and chemical intuition.
Chemical intuition, however, introduces a human bias that might prevent the discovery of the exact molecular constitution by not considering specific isomers if they do not seem probable based on the chemist's experience. This bias might constitute a hurdle in finding novel, undiscovered constitutional isomers. 
In view of the large number of constitutional isomers of medium sized molecules, the probability of picking the right structure from the first estimate is small. Moreover, chemical intuition requires manual intervention, limiting the number of potential target compounds that can be identified.

Thus, it is advisable to support proposed structures by comparison of as many physically measurable properties as available to gain confidence in the correctness of the chosen structure or rule out structures not consistent with experimental observations. Due to the low concentrations of compounds in gas phase experiments, spectroscopic methods that provide conclusive information about the chemical constitution, such as NMR, are unfortunately not applicable.
The high absorption coefficients of BrC, however, allow the use of UV/Vis spectroscopy to probe if the proposed structure is consistent with the experimental observations. Furthermore, it is questionable if only one constitutional isomer is present. Given the chemical complexity of SOA, it is often more realistic to assume a composition of different isomers with similar physical chemical properties.

\begin{figure}
    \centering
    \includegraphics[scale=0.15]{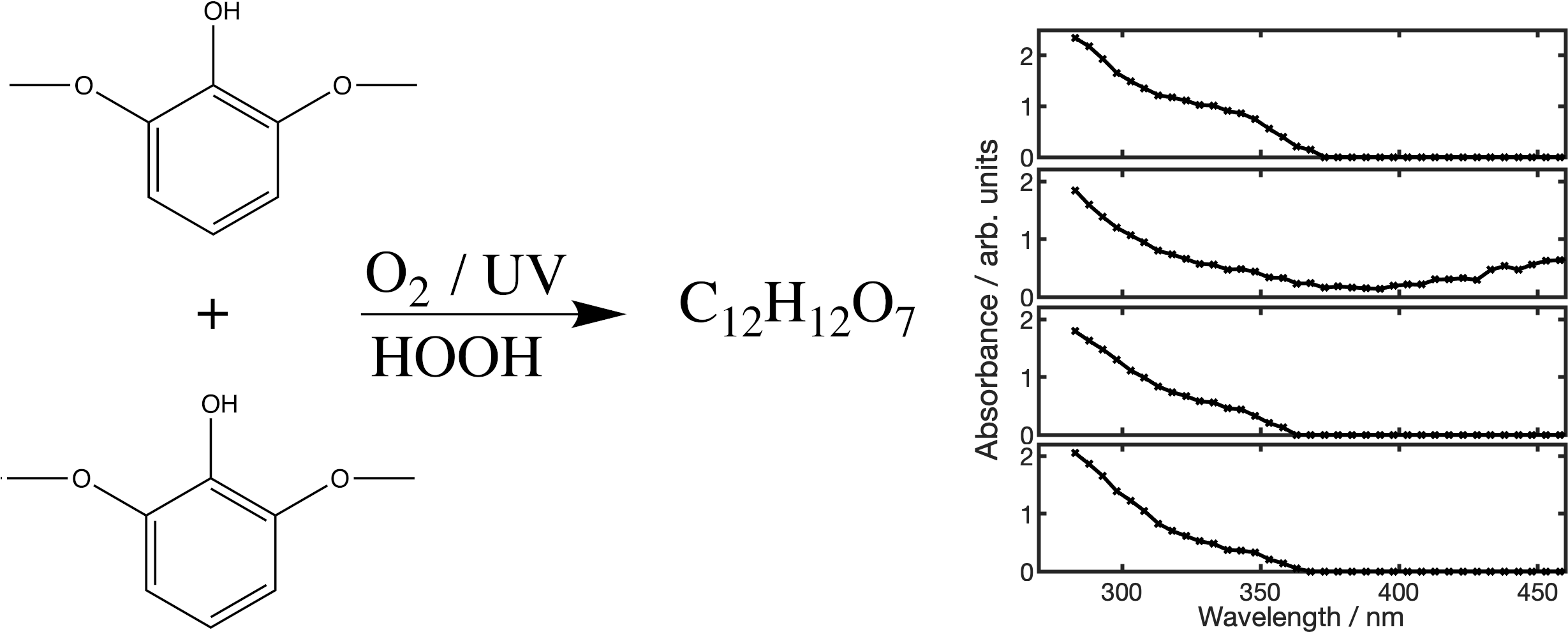}
    \caption{In the proposed reaction scheme (left), aqueous-phase syringol photooxidation forms C$_{12}$H$_{12}$O$_7$, a product with unknown structure that correlates with brown carbon formation. Right: four different UV/Vis absorbance spectra, measured in a filter extract from the CESAM chamber at four different retention times; each corresponds with elution of a different C$_{12}$H$_{12}$O$_7$ isomer.}
    \label{fig:4_spectra}
\end{figure}

In an attempt to rule out any human bias in the proposition of candidate structures, we attack the problem of finding consistent isomers by initially considering {\it all} possible isomers exhaustively; this is in stark contrast to the common approaches based on chemical intuition.\cite{yu2014chemical,de2018nitrogen}
However, due to the quickly growing size of the chemical space with the number of atoms, this approach is already challenging for small and medium-sized molecules and becomes impossible for larger molecules. As a test case, we consider syringol (2,6-dimethoxyphenol), an aromatic phenolic C$_8$ compound (Fig.~\ref{fig:4_spectra}) that has been used as a marker for wood smoke emissions in the atmosphere.\cite{LAURAGUAIS201243} 
When syringol is photooxidized with OH radicals or triplet carbon ($^3$C$^*$) species in the aqueous phase, one of the seven major products detected by negative-mode nano-desorption electrospray mass spectrometry has the sum formula C$_{12}$H$_{12}$O$_7$.\cite{yu2014chemical}  In experiments at the CESAM chamber\cite{wang2011design} where syringol was oxidized with OH radicals, product peaks identified by UHPLC-(+)ESI-MS in aerosol extracts were categorized by whether their concentrations were higher in experiments where brown carbon formed.  Within this group of peaks that correlated with brown carbon, 2\% of the peak area was due to C$_6$ (monomer) products, 94\% was due to C$_{10}$ - C$_{15}$ (dimer) products, and 4\% to larger products, up to C$_{29}$.  Among molecules with less than 20 heavy atoms, C$_{12}$H$_{12}$O$_7$ was the largest peak, responsible for 7\% of the peak area correlating with brown carbon, and co-eluting with an absorbance peak. 
Thus, C$_{12}$H$_{12}$O$_7$ is an appropriate brown carbon candidate (Fig.~\ref{fig:4_spectra}).
Considering all possible molecular graphs (i.e. the set of all atoms and their bonds including bond orders) of C$_{12}$H$_{12}$O$_7$, we assign one graph node for each heavy atom. In total there are more than $10^{35}$  simple connected graphs of 19 nodes\cite{OEIS001349}. The number of molecular graphs is even higher, since this count neither includes elemental composition nor bond orders. 
These large numbers show that the major bottleneck in exploring this chemical subspace lies in the efficiency of the computer generation of molecular structures, which ultimately limits this approach for molecules larger than a given size.
A second challenge arises from the prediction of physical chemical data for all these structures needed to determine the candidate molecules consistent with experimental measurements. In BrC, the observable usually is the UV/Vis spectrum, which can be predicted reasonably well by correlated quantum chemistry methods.\cite{send2011assessing,Cisneros2017,Tapavicza2018,tapavicza2019generating,loos2020mountaineering} However, the computational resources necessary for the spectra prediction of such a large number of isomers quickly becomes out of reach.  

Here, we developed a computational workflow to find possible constitutional C$_{12}$H$_{12}$O$_7$ isomers consistent with the recorded absorption spectra.   
To tackle the exhaustive generation of constitutional isomers we present a graph-based, bias-free molecule generator, that leverages massively parallel computation. The problem of quantum chemical spectra prediction of a large number of molecules is solved by making use of machine learning to predict spectral properties of the molecules.\cite{ramakrishnan2015electronic}
In a Monte-Carlo procedure, we then determine the likelihood that specific feature groups give rise to the experimentally observed spectrum.

The work flow  starts from an unbiased and exhaustive generation of all possible molecular graphs. The number of graphs is further reduced by molecular stability and steric criteria based on tight-binding density functional theory. After prediction of electronic excitation energies and oscillator strengths, we filter the compounds by the probability of agreement with experimental UV/Vis absorption spectrum.  Finally, we explore how additional information about structure or functional groups could further reduce the number of possible C$_{12}$H$_{12}$O$_7$ isomers consistent with experimental data.

\section{Materials and Methods}
\subsection{Experimental}
The filter extraction protocol has been described previously\cite{Wang2017chemical}. Briefly, each collected Teflon filter (\( 1\ \mu \)m pores, 47 mm diam.) was spiked with caffeine (final concentration 100 ppb), as internal standard and then extracted twice with 6 mL of acetonitrile and agitated for 20 minutes with an orbital shaker at 1000 rpm. The extracts were then filtered with a syringe filter (0.2 $\mu$m, Pall Acrodisc® PSF, with GHP membrane, hydrophilic polypropylene) to remove any insoluble particles and blown dry under a gentle N$_{2}$ (g) stream at ambient temperature. The residues were reconstituted in 0.2 mL of water:methanol (v/v 1:1, Optima®LC/MS, Fischer Scientific). Finally, the filter extracts were analyzed by ultra-high performance liquid chromatography (Dionex 3000, Thermo Scientific) using a Water Acquity HSS C18 column (\( 1.8\ \mu \)m, 100 x 2.1mm) coupled with a diode array UV/Vis absorbance detector and a Q-Exactive Hybrid Quadrupole-Orbitrap mass spectrometer (Thermo Scientific) equipped with an electrospray ionization (ESI) source operated in positive or negative mode. The mobile phase used was constituted of (A) 0.1\% formic acid in water (Optima® LC/MS, Fischer Scientific) and (B) 0.1\% formic acid in acetonitrile (Optima® LC/MS, Fischer Scientific). Gradient elution was carried out by the A/B mixture at a total flow rate of 300 $\mu$L/min:
0 to 13 min B from 1\% to 100\% , 13.1 min B 1\% for 9 min.

Raw data was processed with MZmine 2.51.  Features with a higher intensity than 1$\times 10^{6}$ and at least 10 times higher than the blank intensity were selected. The chromatographic peaks of all ID selected were visually analyzed and a proposed molecular formula was obtained using Xcalibur  2.2 (Thermo Scientific) software package.
A subset of peaks was then identified that had areas averaging at least 5 times larger in experiments where brown carbon formed than when it did not; C$_{12}$H$_{12}$O$_7$ was prominent within this subset.  

\subsection{Molecule generation}
For the sum formula C$_{12}$H$_{12}$O$_7$, we systematically\cite{McKay201494} enumerate all molecules that potentially could be a product of the reaction in the atmospheric chamber. We rationalize that the product forms via radical-initiated coupling\cite{chang2010characterization} of two syringol units to C$_{16}$H$_{18}$O$_6$, the most abundant SOA product identified in previous studies,\cite{sun2010insights,yu2014chemical} followed by further oxidation and fragmentation to C$_{12}$H$_{12}$O$_7$.\cite{yu2014chemical} 
We limited ourselves to those candidates where the two C$_6$-rings found in the two reactant molecules persist in the product, which requires the loss of methoxy carbons.
We note that half of the syringol SOA product structures proposed by Yu et al.\cite{yu2014chemical} have lost at least one methoxy carbon, and 16\% of their proposed structures have lost all methoxy carbons.  Demethylation of methoxy groups during photooxidation has been observed for vanillin,\cite{vione2019formation} syringaldeyde, and acetosyringol.\cite{huang2018formation}  
Technically, this enumeration is performed by a) enumerating all potential molecular graphs ignoring hydrogens, b) constructing all possible hydrogen saturations of these graphs, c) filtering all molecules which are not stable in GFN2-xTB calculations\cite{bannwarth2019gfn2}. The protocol for these steps is detailed in the SI and is based on Refs.~\cite{McKay201494,1323804,o2011open,openbabel,halgren1996merck}.

\subsection{Electronic structure methods and machine learning}


To assess the absorption spectrum, we computed the lowest three excitation energies and their corresponding oscillator strengths using the Algebraic Diagramatic Construction to Second Order (ADC(2)) method.\cite{schirmer1982beyond,hattig00} To include effects of water solvation in  the  calculation,  we  employed  the  Conductor-like  Screening  Model  (COSMO)\cite{Klamt1993} using a dielectric constant of 80.1 and a refractive index of 1.3325.\cite{lunkenheimer2013solvent} The def2-TZVP basis set was used.\cite{Weigend2005} This approach has been shown to yield accurate excitation energies.\cite{Thompson2018} Calculations were carried out with TURBOMOLE V7.2.\cite{TURBOMOLE, balasubramani2020turbomole}


Since it is prohibitively expensive to apply this reliable method to the exhaustive list of all molecules, we calculated 10,000 randomly selected molecules as training set for the Kernel-Ridge-Regression (KRR) method\cite{Rupp2012} with the FCHL molecular representation\cite{faber2018alchemical} as implemented in the QML toolkit \cite{QML}. Machine learning in general and KRR in particular have been successfully used to predict excited state properties\cite{hase2019machine,westermayr2020machine,xue2020machine}, typically highlighting the need for high-quality reference data. 82 molecules were excluded since they exhibited negative excitation energies, which indicates a non-stable ground state. We determined optimal hyperparameters for the kernel widths and regularizer with 5-fold cross validation (see SI).
Once both the excitation energies and oscillator strengths for the lowest three excitations have been predicted for all compounds from machine learning, we can model the spectrum\cite{Schalk2016,Epstein2013} and compare it to the experimental ones. We employ a Monto-Carlo method (see SI) to assess whether these predicted spectra are compatible with the experimental spectra. In this work, a predicted spectrum is considered compatible if the experimental spectrum and predicted spectrum are separated by at most one standard deviation of both modeling and experimental uncertainties.

\section{Results and discussion}
\subsection{Analysis of the generated molecules}
\begin{figure}
    \setlength{\lineskip}{0pt}
    \centering
    \includegraphics[width=\textwidth]{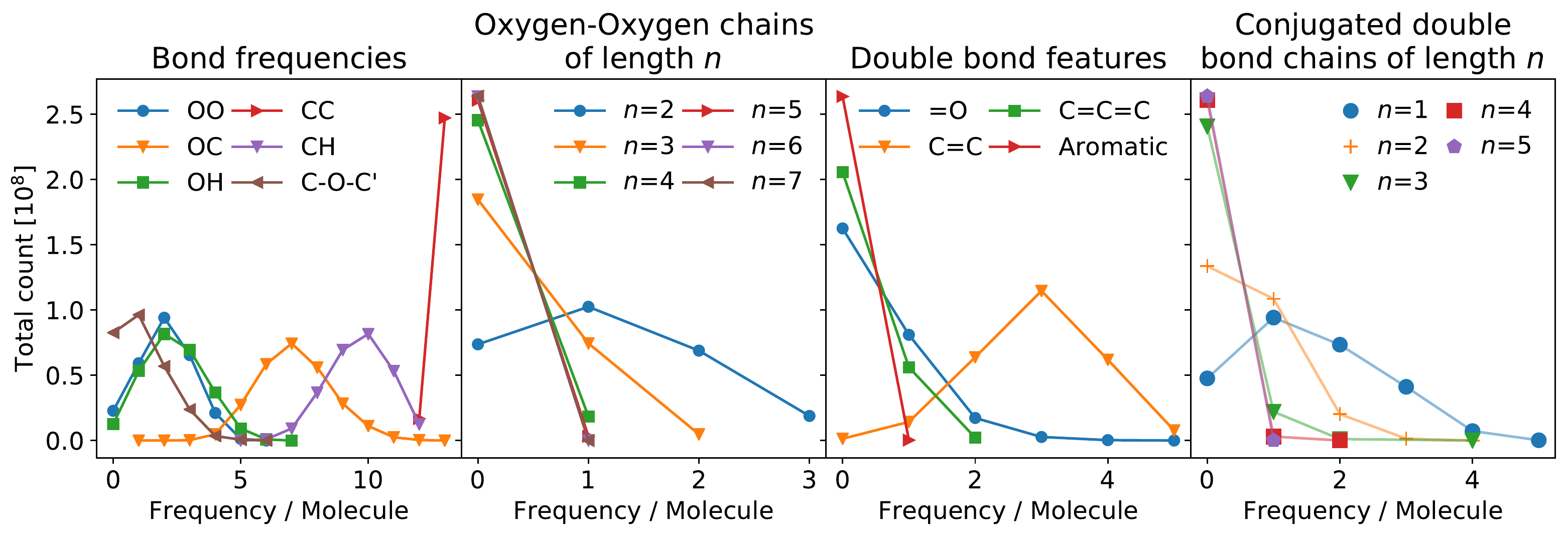}
    \caption{Total number of molecular graphs with given feature occurrences. Panel 1: bond count frequencies, and ether bridges connecting the two carbon rings; panel 2: count of oxygen-oxygen chains of length $n$; panel 3: count of carbonyl sites, allene sites, double bonds, aromatic rings; and panel 4: conjugated double bond chains of length $n$. Each curve adds up to the total number of molecular graphs of 263'917'411. }
    \label{fig:oo}
\end{figure}

At first we will analyze the distribution of features in the molecular graphs, before the structures have been optimized.
According to our initial assumption that two C$_6$-rings exist in the structure, there are two different possibilities how the rings are connected: either directly by a carbon-carbon bond, or by one or more oxygen atoms, serving as a bridging unit. These two possibilities are reflected by having either 13 or 12 C-C bonds, respectively (Fig.~\ref{fig:oo}). Analyzing C-O bonds, we find a peaked distribution ranging from 1 to 13, with a maximum probability at 7. Oxygen-oxygen bonds range from 0 to a maximum of 6, with the maximum of 6 corresponding to a structure where an O$_7$ chain exists (blue, middle Fig.~\ref{fig:oo}). We also note that the longer the oxygen chain, the fewer graphs are found, as expected. We note that most structures have 0-2 carbonyl groups (Fig.~\ref{fig:oo}, right), which are important for absorption properties.
 
Of the 263 million graphs about 123 million lead to stable three-dimensional structures according to GFN2-xTB. 
All their coordinates are available online\cite{datasetgeo} together with the reference data for the machine learning model\cite{datasetgeoener}.
Since we are mainly interested in the structures that are consistent with the experimental spectra, we skip a more detailed analysis of the features of this large structure set. However, it is important to say that we observe a substantial amount of structures that are not commonly seen. For instance, we find a considerable number of stable molecules with chains up to seven oxygen atoms and dioxiranes (i.e. three rings with two oxygen atoms).
There has been an ongoing discussion about the possible length of oxygen chains \cite{mckay1998long}. While it might seem unlikely to find oxygen chains with more than three members, theory has predicted the stability of oxygen chains up to at least 6 members. Experimentally, four-membered chains have been confirmed\cite{mckay1998long}.
Dioxiranes have been known experimentally since 1978, although their existence were already predicted in 1899 by Bayer and Villiger\cite{rappoport2007chemistry}. 
An indication of the relative stability of the molecules can also be based on total electronic energies (see Fig. 1, SI).

\subsection{Electronic spectra prediction}
\begin{figure}
    \centering
    \includegraphics[scale=0.3]{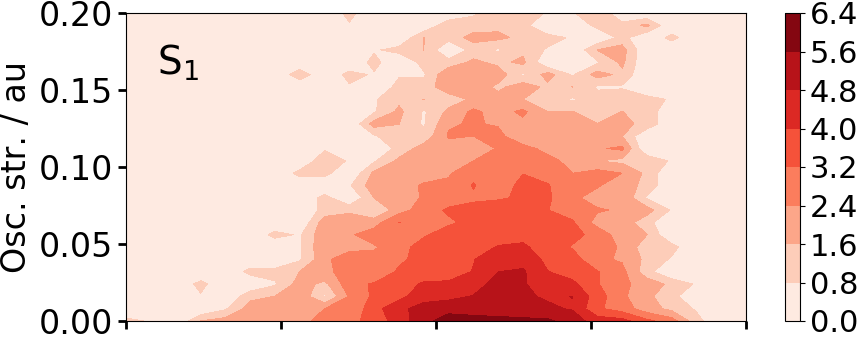}\\
    \includegraphics[scale=0.3]{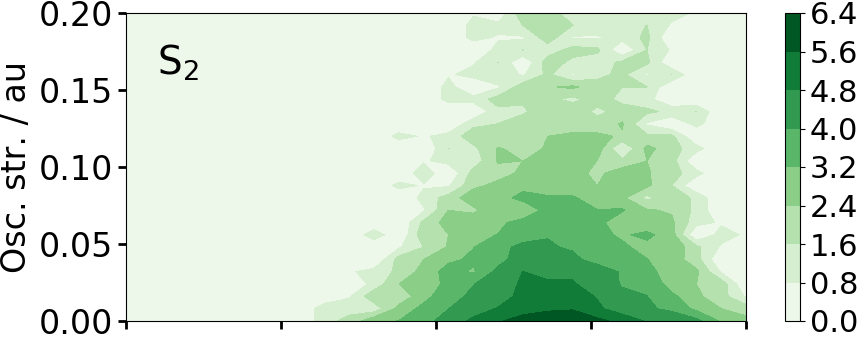}\\
    \includegraphics[scale=0.3]{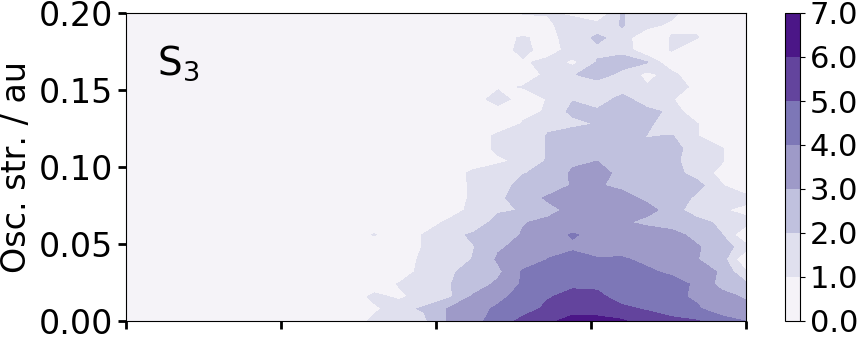}\\
    \hspace*{0.1cm}\includegraphics[scale=0.3]{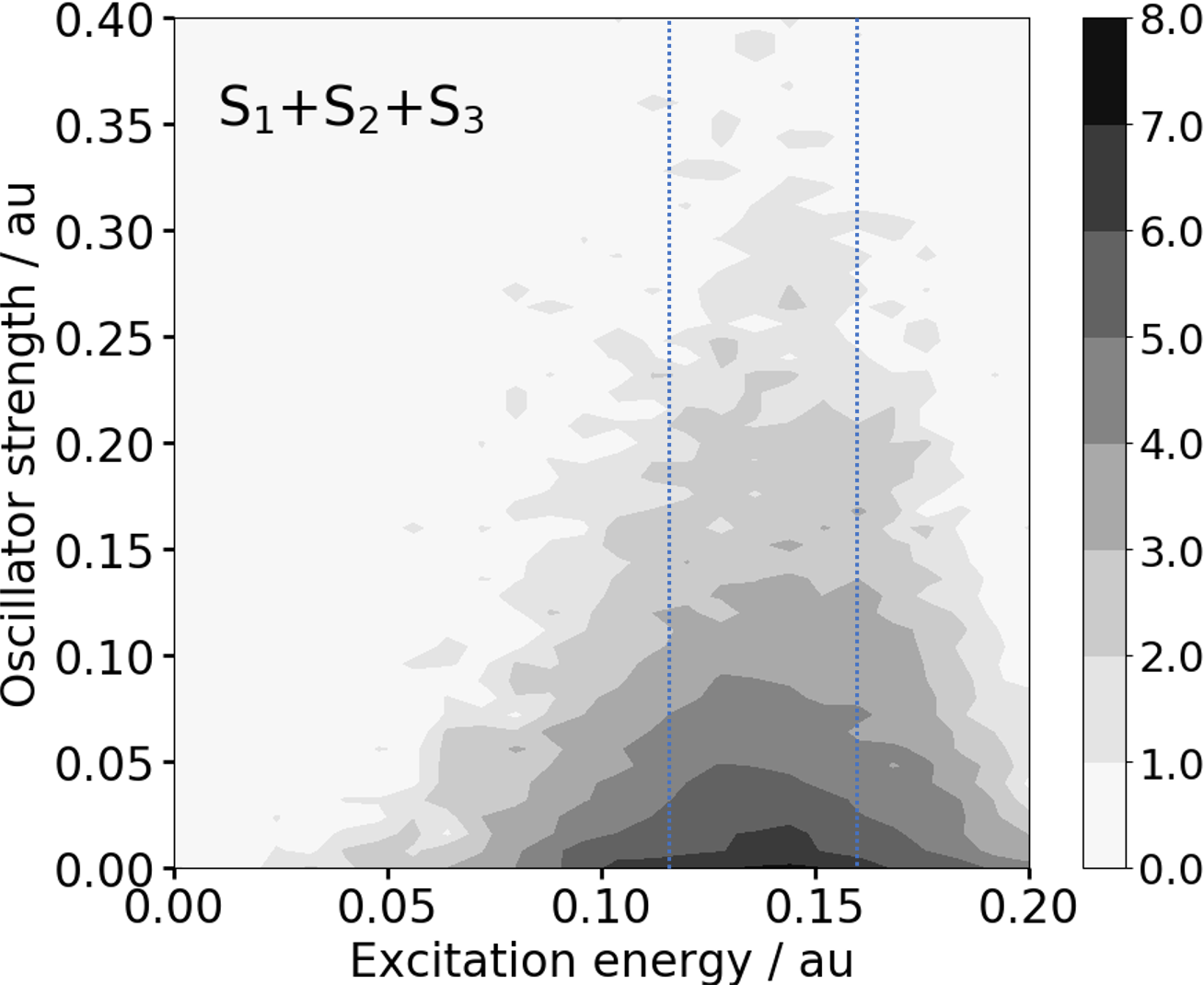}\\
    \caption{Distribution of ADC(2)/COSMO excited states as a function of excitation energy and oscillator strengths of the 9,918 training molecules. The distribution is shown separately for the lowest three excited states (top three panels) and combined for all three states (bottom panel). The color code refers to the decadic logarithm of the density found in a square of an area of 0.008$\times$0.008 au$^2$. The blue dotted lines in the bottom panel indicate the region in which the experimental band is located.}
    \label{fig:exci_dens}
\end{figure}

\begin{figure}
    \centering
    \includegraphics[scale=0.4]{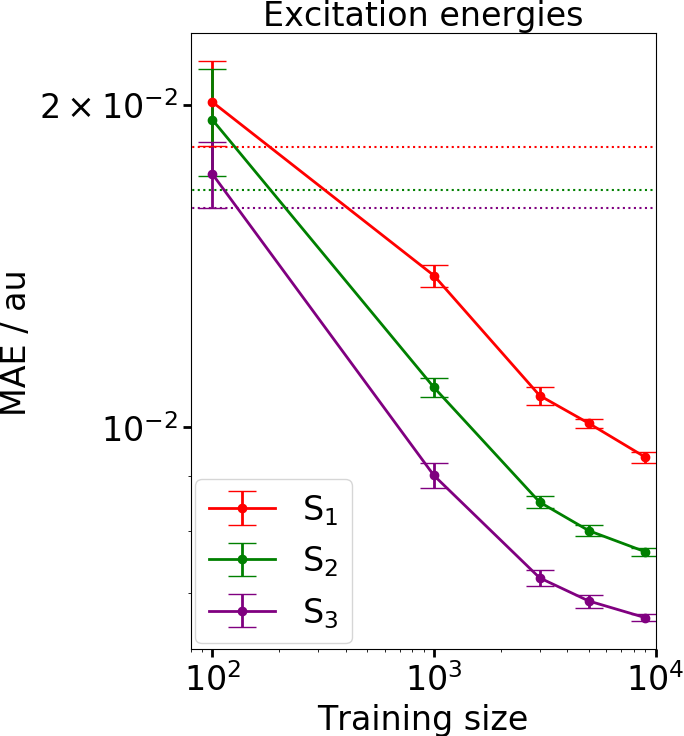}
    \includegraphics[scale=0.4]{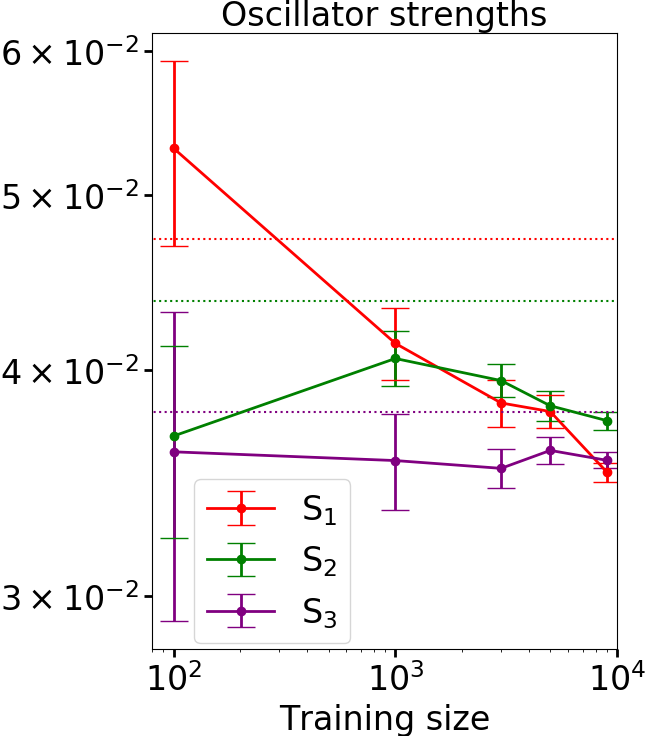}
    \caption{Left: Mean absolute error (MAE) as a function of training set size for the learning of the lowest three excitation energies. Right: Mean absolute error (MAE) as a function of training size for the learning of the lowest three oscillator strengths. In both plots, the dashed lines indicate the error of the null model, using the same color code for the different states.}
    \label{fig:learning_syringol}
\end{figure}

For a random set of 9,918 molecules ADC(2)/COSMO calculations resulted in positive excitation energies and oscillator strengths for the lowest three states (Fig.~\ref{fig:exci_dens}). The data is available online\cite{datasetgeoener}.
From Fig.~\ref{fig:exci_dens}, we see that the first excited state contributes with the highest oscillator strengths in the region between 0.12 and 0.16 au, where the experimental absorption band is located, but S$_2$ and S$_3$ also show substantial absorption in this region.
Using KKR, we predicted the lowest three excitation energies and oscillator strengths based on different training set sizes
(Fig.~\ref{fig:learning_syringol}). For a training set of 9000 molecules, predictions exhibit mean absolute errors (MAEs) of 9, 8, and 7~mHa, for S$_1$, S$_2$, and S$_3$, respectively.
Thus, ML errors are similar to the expected error of ADC(2) with respect to experimental values, 
which was previously determined to be 8~mHa (0.21~eV)\cite{sarkar2020benchmarking}. 
The curves confirm the learning abilities of the model as it makes use of additional training data to improve prediction accuracy. The accuracy of the machine learning predictions is set into perspective by comparison to the null model (dashed lines in Fig.~\ref{fig:learning_syringol}),  which is obtained when the mean excitation energy over all training molecules is used as prediction.
MAEs for the oscillator strengths amount to 0.035, 0.038, and 0.036 au, for S$_1$, S$_2$, and S$_3$, respectively (Fig.~\ref{fig:learning_syringol}, right). Interestingly, oscillator strengths of S$_1$ benefit the most from KKR, whereas, for S$_2$ and S$_3$, learning curves are comparably flat.
Using the ML model based on the 9,918 training molecules, we predicted the lowest three excitation energies and oscillator strengths for the remaining 120 million stable structures.
\subsection{Establishing matching characteristics}
\begin{figure}
    \centering
    \includegraphics[width=\textwidth]{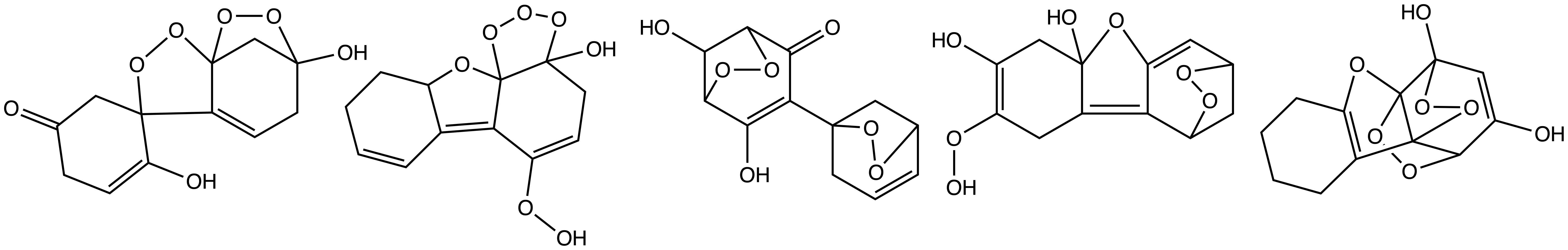}
    \includegraphics[width=\textwidth]{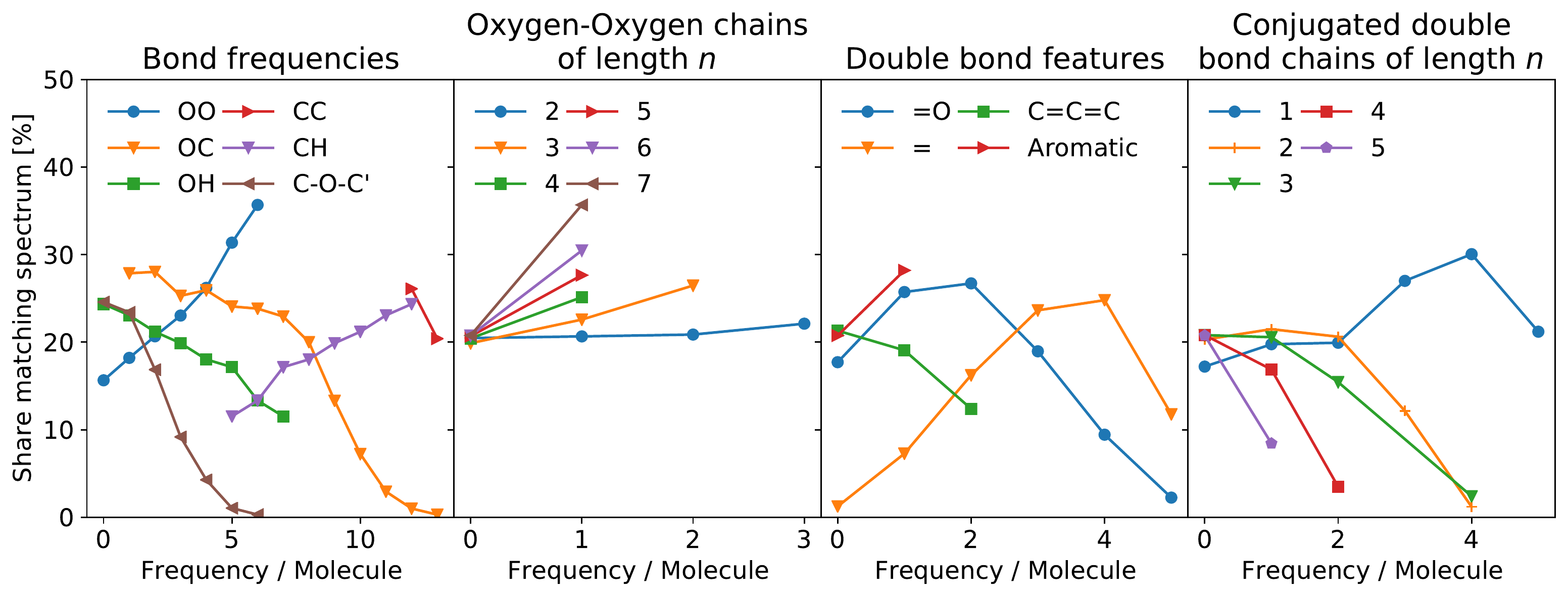}
    \includegraphics[width=\textwidth]{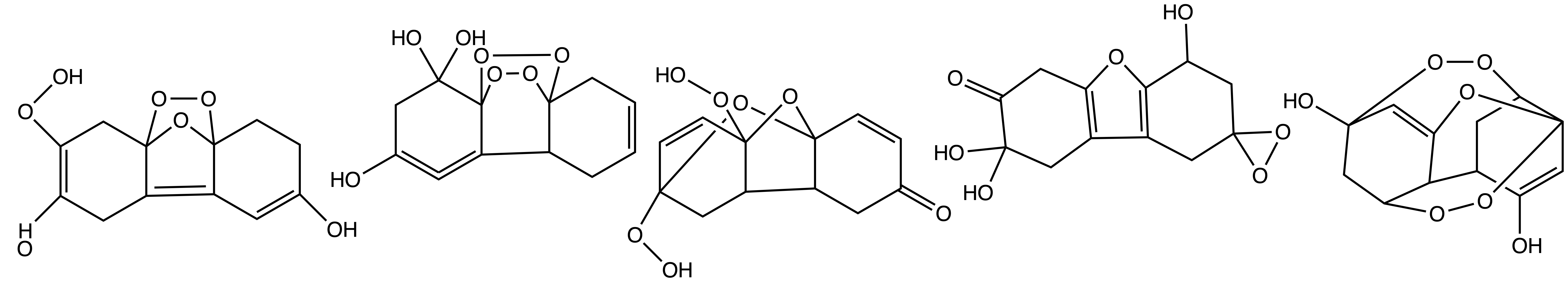}
    \caption{Per-feature probability (Share matching spectrum) of molecular structures being compatible with the first spectrum in Figure~\ref{fig:4_spectra} shown in the panels.
    Conditional probabilities are exemplified by grouping all molecules by their feature vector and calculating the share of matching molecules for each group. Low-energy representatives of the largest groups are shown for matching (top) and non-matching (bottom) molecules.}
    \label{fig:features}
\end{figure}

We present the analysis for the first spectrum on the top right of Fig.~\ref{fig:4_spectra}; results for the remaining three spectra are very similar (see Fig. 2, SI). Out of the 123 million stable molecules, 55 million match this spectrum according to the criteria defined in the SI.
For every structure, we determined a feature vector that describes the structural features in the molecules (Fig.~\ref{fig:features}). Features considered were: a) bond types, b) oxygen chains of different lengths, c) carbonylic groups, d) double bonds, e) conjugated double bonds, f) aromatic rings, g) ether bridges, and h) allene groups.
The total dimension of the feature vector amounts to 21, whereas the length of the entries vary between 2 and 6. 
For instance for the carbon-carbon bonds, only two values are possible (12 and 13), but for two-membered oxygen chains the number of entries amounts to four, because possible values are 0, 1, 2, 3. 
For every feature and for every number value thereof, we calculate the fraction of the molecules that are compatible with the experimental spectrum as defined above. This allows to correlate molecular features with the probability that it causes the experimentally observed spectrum (see Fig.~\ref{fig:4_spectra}). Analyzing the features in Fig.~\ref{fig:features}, we find that for example the probability that a molecule is consistent with experimental spectrum increases with the number of (O-O) bonds (blue line, left panel).
As another example (orange line, right panel), we see that the probability of a matching molecule decreases if there are more than two cases of conjugated double bond chains of length two.

To illustrate the chemical diversity of stable molecular structures that are compatible with the experimentally observed spectrum, we group all molecules by their feature vector. Representatives of large feature groups of matching and non-matching molecules are given on top and bottom of Fig.~\ref{fig:features}, respectively. The corresponding groups of molecules are huge: just for the first molecule on the top left in Fig.~\ref{fig:features}, there are 695,039 stable molecules that match the experimental spectrum and have an identical feature vector.

Each of the other molecules shown in Fig.~\ref{fig:features} is just one representative of similarly large groups of feature-identical stable molecules.
While the presence of individual molecular features can significantly reduce the number of molecules, the sheer size of these groups highlights that the share of molecules matching any spectrum is still by far too large to claim unique identification.
Thus, the extent of the structural ambiguity of brown carbon absorption spectra is made clear by the exhaustive enumeration of all possible molecular structures.

\subsection{Filtering strategies}

In view of these large numbers of candidate structures, it is evident that any identification of individual molecules based on their spectra needs more criteria derived from experiments to reduce the number of possible candidates. 
In practice, misidentifications are likely if too few additional constraints are included in the search.
Furthermore, a comparison between the representative matching and non-matching feature groups (see Fig.~\ref{fig:features}) shows that it is not trivial to establish obvious structural characteristics that would increase the likelihood of being consistent with the experimental spectrum.
Hence, common textbook relationships between structural elements and absorption properties (e.g.\ batochromic shift) are of limited utility in the selection of candidate brown carbon molecules.

\begin{table}[h]
    \centering
    \begin{tabular}{l|r}
    Total molecules with two C$_6$ rings & 263,917,411\\
    and which have OH groups & 263,917,411\\
    and which have no oxygen chain longer than 2 & 161,160,394\\
    and which have an oxygen connecting the carbon rings & 115,715,458\\
    and which have one aromatic ring & 134,944\\
    and which are stable & 64,121\\
    and which match spectrum 1 & 36,518\\
    \end{tabular}
    \caption{Summary of how given structural features reduce the number of possible C$_{12}$H$_{12}$O$_7$ structures.} 
    \label{tab:cond_prob}
\end{table}

Strategies to obtain more decisive criteria in establishing the possible candidates can be based on structural motifs found in MS fragmentation data, MS ionization data, and/or stability criteria. Applied to our first spectrum, Table~\ref{tab:cond_prob} lists how these criteria reduce the number of possible structures. 
In the present case, although fragmentation spectra of the individual C$_{12}$H$_{12}$O$_7$ isomers are not available, the detection of both hydrogen and sodium ion adducts of the C$_{12}$H$_{12}$O$_7$ isomer in question suggests that it contains OH and ether groups rather than carbonyls.\cite{Swanson2017}  Furthermore, if we exclude oxygen chains longer than two (which most likely are not stable enough to endure the analytical procedure), only 36'516 stable molecules are left that match the experimental spectrum. This constitutes 0.01 and 0.03 \%  of the initially generated molecular graphs and stable structures, respectively. Given sufficiently accurate computational chemistry methods, the total energies of the structures (Fig. 1, SI) could be used to select or exclude certain structures; due to the approximative character of the GFN2-xTB calculations, we do not pursue this route further.

Starting from a complete list of all molecules is key to allow a bias-free filtering based on experimental input. Most importantly, filtering molecular graphs by MS fragments (or electrospray ionization information) is free of approximations from the theory side as no filtering based on computational chemistry or machine-learning methods is done at these early stages. The presence or absence of a structural feature in a given molecular graph can be determined readily. 

Having a substantially shortened list (e.g. the 0.01\,\% for our case) allows for better calculations on the theory side once the filtering possibilities based on MS data are exhausted. There are two reasons for this: not only does a smaller chemical space require fewer training points to be accurately modeled with a machine learning approach, but also the reference data for the individual training points can be calculated using a better level of theory with fewer approximations. 

\begin{figure}
    \centering
    \includegraphics[width=0.45\textwidth]{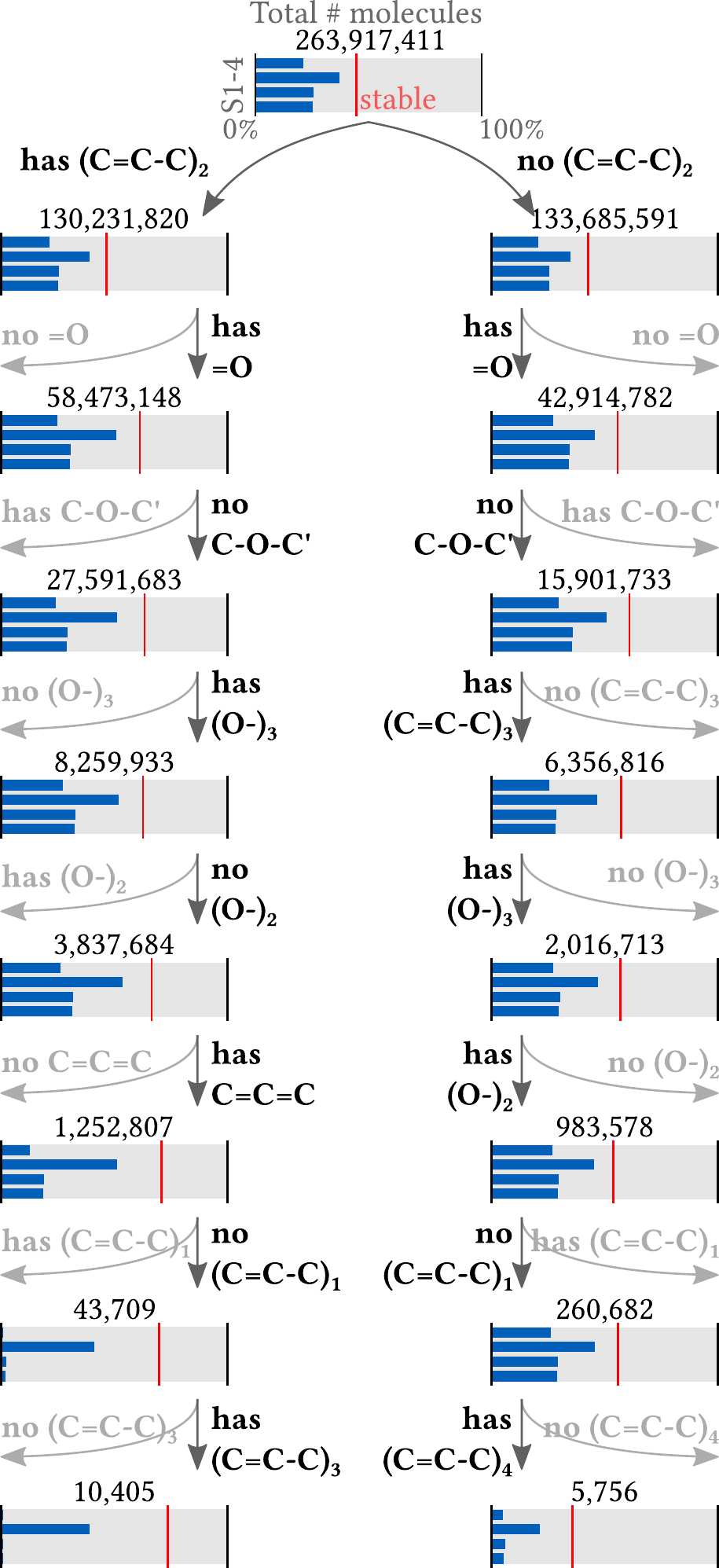}
    \caption{Idealized reduction in number of molecules as more and more conditions on the molecular graph are applied (from top to bottom). Note that these conditions are not founded in experimental fragmentation data, but rather illustrate the filtering process. Including all features would give a wide tree, so only two branches of the tree are shown. After each filter step, the total number of remaining molecules is shown where the red bar denotes the share thereof that is stable. The four bars illustrate how many of the residual molecules are compatible with the spectra 1-4 in this work. (X)$_n$ denotes that the structural feature X appears $n$-times in a row.}
    \label{fig:tree}
\end{figure}

Figure~\ref{fig:tree} illustrates how this filtering could be employed in a systematic fashion by repeatedly searching for structural features 
that divide the current set into two new sets of as equal size as possible. Similar to the method of binary search, this filters the total list of molecules in the fastest possible way if only tests for the existence of particular MS fragments are allowed. 
For the chemical space under consideration, an average of 15 fragment tests would be required to reduce the number of candidate molecules to below 10,000, if fragments were randomly distributed amongst all stable molecules and independent of each other. Typically, this is not the case, as exemplified by Figure~\ref{fig:tree} where we require tests for eight fragments until the number of molecules has been reduced to about 10,000 
which would then be accessible for quantum chemistry calculations. In practice, this means that typically on the order of ten fragment tests would be needed to narrow the molecules down. For larger molecules, and particularly for molecules with a potentially branched structure, 
the number of fragments tests required will be larger. Starting from the exhaustive list of molecules however, it is clear exactly how many molecules remain to be analysed and thus going down such a decision tree could guide experimental work or MS data analysis. 
Furthermore, such an approach can not only determine whether additional criteria are still needed to identify a molecule, but can also identify which criteria will most efficiently narrow down the candidate molecule list.

The information whether molecules with these features are stable is typically not available while filtering the molecules, as long as the list of molecules is too long to render the required calculations feasible for multiple spectra. In this work however, we have performed the stability calculations for the complete list to illustrate in Figure~\ref{fig:tree} that the structural features alone are not always sufficient to determine stability or similarity to a UV/Vis spectrum. As the number of fragmentation results included increases in Figure~\ref{fig:tree}, the share of stable molecules and those matching the four spectra in this work are initially roughly constant along the two paths shown. Only at the final stages does the feature list become more sensitive to the spectra in question. This emphasizes that real-world structure determinations will typically require a substantial number of confirmed/missing MS fragment determinations in addition to the UV/vis spectrum.



We have systematically enumerated all molecules with the sum formula C$_{12}$H$_{12}$O$_7$ containing two C$_6$-rings. We investigated whether the specific C$_{12}$H$_{12}$O$_7$ isomer behind an experimental brown carbon UV/Vis spectrum can be identified uniquely if a bias-free systematic comparison is done. To this end, we used a machine learning model to predict spectra for all possible 123 million stable molecules in the set. We find that the experimental spectrum alone only halves the set of possible candidate molecules, so much additional information is required to determine the structure of a brown carbon molecule. Even with multiple MS fragments identified, there are tens to hundreds of thousands of potential structures that are compatible with the spectrum. 
The true scale of this problem only becomes clear once the exhaustive enumeration is done.

In light of our findings, we still consider identifying functional groups from MS the most promising strategy to reduce the number of candidates, especially if this information can be used early during the generation of molecular graphs. The advantage of using this information early is that it can be used to accelerate the graph generation. In addition, it reduces the chemical diversity, which may then reduce the error of the machine learning model.

Without the systematic enumeration of molecular targets, it becomes unclear whether sufficiently numerous molecular fragments have been identified to narrow down the list of potential molecules. This might lead to mis-identifications of molecules: Laskin et al.\cite{yu2014chemical} suggested a possible structure for a C$_{12}$H$_{12}$O$_7$ product found in a syringol photooxidation chamber study, but our calculation shows that because of its dominant absorption band between 350 and 400~nm, the spectrum of this structure (Fig. 4, SI) is not consistent with any of the four experimentally measured spectra shown in Figure 1.

Based on the numerical evidence in this work, we expect that a systematic enumeration approach, where high-quality MS fragmentation data is included early on and where calculated spectra come from machine learning predictions based on quantum chemistry calculations, will make possible the rapid identification of individual brown carbon molecules based on their exact mass, MS fragmentation spectrum, and UV/Vis spectrum.  In addition, such an approach will also yield guarantees that there are no other molecules that also would fit the experimental data.




\begin{acknowledgement}
We would like to thank Stefan Heinen and Anders S. Christensen for support with the QML code.
Research reported in this paper was supported by National Institute of General Medical Sciences of the National Institutes of Health (NIH) under award numbers UL1GM118979-02, TL4GM118980, and RL5GM118978 and NSF award number AGS-1826593. The content is solely the responsibility of the authors and does not necessarily represent the official views of the NIH. We acknowledge technical support from the Division of Information Technology of CSULB.
O.A.v.L. acknowledges support from the Swiss National Science foundation (407540\_167186 NFP 75 Big Data) and from the European Research Council (ERC-CoG grant QML and H2020 projects BIG-MAP and TREX). 
This project has received funding from the European Union's Horizon 2020 research and
innovation programme under Grant Agreements \#952165
and \#957189.
This result only reflects the
author's view and the EU is not responsible for any use that may be made of the
information it contains.
This work was partly supported by the NCCR MARVEL, funded by the Swiss National Science Foundation. 

\end{acknowledgement}

\begin{suppinfo}

\begin{itemize}
    \item Histogram of total ground state energies of 2 \% of randomly selected structures.
    \item Correlation of features with compatibility of spectra 2 --4.
    \item Figure with tolerance regions of the spectra and illustration of matching probability.
    \item Description of procedure to generate molecules.
    \item Computational details of machine learning procedure.
    \item Description of Monte-Carlo procedure to determine matching probability with experimental spectra.
    \item Computational results of the proposed structure of Laskin et al.
\end{itemize}

\end{suppinfo}

\bibliography{papers,misc}

\end{document}